# Resonant transmission suppression in high-index nanoparticle arrays

Viktoriia E. Babicheva[1,*] and Andrey B. Evlyukhin[2,3]

[1]College of Optical Sciences, University of Arizona, 1630 E. University Blvd., P.O. Box 210094, Tucson, AZ 85721
[2]Moscow Institute of Physics and Technology, 9 Institutsky Lane, Dolgoprudny 141700, Russia
[3]ITMO University, 49 Kronverksky Ave., St. Petersburg, 197101, Russia
*email: vbab.dtu@gmail.com

**Abstract.** High-index nanoparticle lattices have attracted a lot of interest recently as they support both optical electric and magnetic resonances and can serve as functional metasurfaces. Here we demonstrate that under particular conditions, the all-dielectric nanoparticle metasurfaces can resonantly suppress transmission. Electric and magnetic dipole resonances of silicon nanoparticle arrays are studied in the air and in the dielectric matrix in visible and near-infrared spectral ranges. We show that the wave resonantly scattered forward by the one or both electric and magnetic dipole moments of nanoparticles can destructively interfere with the incident wave, providing significant suppression of the transmission through the array. The reported effect can find important applications in different fields related to optics and photonics such as the development of filters, sensors, and solar cell.

Being a few hundred nanometers in size, nanoparticles with high refractive index and low absorption losses, such as silicon, support strong electric and magnetic resonances in optical range and hold a lot of promise for future photonic devices [1-4]. A variety of effects related to the interplay of electric and magnetic resonances have been studied, and large attention has been paid to directional scattering and Kerker effect [5-10]. Furthermore, arranging particles in the array brings new optical phenomena. For instance, when the array period is comparable with the wavelength of resonances, lattice resonances (LR) can be excited in the proximity to the Rayleigh anomalies, where a transition from a diffracted order propagating in the plane of the array to an evanescent diffracted order is realized [11]. It was shown in [1] that, in the case of Si nanoparticles, these resonances can correspond to the electromagnetic coupling between the nanoparticles in arrays due to resonant excitations of their electric and magnetic dipole moments. By arranging nanoparticles in the lattice, one can achieve very narrow resonances, and they can be used for sensing along with their plasmonic counterparts [12]. The positions of electric- and magnetic-lattice resonances can be controlled by the mutual-perpendicular periodicities of the array. If these resonances are brought into overlap and have comparable magnitudes and phases, the resonant Kerker effect (resonant suppression of light reflection) occurs [13,14]. Combining array with a substrate, one can compensate wave reflected from the arrays itself and from the substrate, which will also cause resonant suppression of the reflection, the so-called substrate-mediated Kerker effect [15].

In the present Letter, we propose and study an additional important resonant effect in the high-refractive-index (silicon) nanoparticle arrays: transmission through such array can be significantly suppressed because of the destructive interference of the incident wave and the wave scattered forward by either electric (ED) or magnetic (MD) dipoles or both. Our study is based on an analytical model with coupled-dipole equations (CDEs) [1,16,17] and numerical simulations with CST Microwave Studio frequency-domain solver. We choose to work with two types of periodic arrays of silicon nanospheres with radius $R$ = 65 nm, as those particles support ED and MD resonances (EDR and MDR, respectively) in the visible range, and with $R$ = 230 nm where particles have the resonances in the near-IR spectral range in the proximity of telecommunication wavelength 1.3-1.6 μm. We study either nanoparticles in the air ($n$ = 1) or in a dielectric matrix with refractive index $n$ = 1.5, and we consider only two-dimensional periodic arrays throughout the work.

Comparison of numerally calculated transmission and reflection spectra of silicon nanosphere array with diameter $d_{sp}$ = 2R =130 nm and a continuous silicon layer of the same thickness $d$ = 130 nm shows that the spectra have the opposite trends: array has transmission minimum where the continuous layer has transmission maximum (Fig. 1). Designing metasurface out of nanoparticle array one can obtain devices with transmission and reflection windows in pre-defined ranges. Figure 1 also includes the transmission and reflection calculated in a framework of the dipole approximation, when every particle of the array is considered as ED and MD [1]. From the comparison between results obtained with CST Microwave Studio and with the dipole approximation, we conclude that the transmission and reflection properties are basically determined by the ED and MD responses of the nanoparticles. This can significantly simplify a physical analysis, explanation, and practical tuning of the array optical properties.

For an infinite periodic two-dimensional array of dipole nanoparticles supporting solely EDR and MDR and located in xy-plane (z = 0), the transmission coefficient for normal light incidence and for non-diffraction spectral regions is determined by the expression [1]

$$t = 1 + \frac{ik_d}{2S_L}\left(\alpha_{\text{eff}}^{\text{E}} + \alpha_{\text{eff}}^{\text{M}}\right), \qquad (1)$$

where $S_L = D_x D_y$ is the area of the array elementary cell, $D_x$ and $D_y$ are dimensions of the cell in x- and y-direction, respectively, $k_d$ is the wavenumber in the medium around the array, $\alpha_{\text{eff}}^{\text{E}} = 1/(\varepsilon_0 \varepsilon_d / \alpha^E - k_d^2 G_{xx}^0)$ and $\alpha_{\text{eff}}^{\text{M}} = 1/(1/\alpha^M - k_d^2 G_{yy}^0)$ are the effective ED and MD polarizabilities of particles in the array, respectively. Here $\varepsilon_0$ and $\varepsilon_d$ are the vacuum dielectric constant and the relative dielectric constant of surrounding medium, respectively, $\alpha^E$ and $\alpha^M$ are the ED and MD polarizabilities of a single particle surrounded by a homogeneous medium, respectively, $G_{xx}^0$ and $G_{yy}^0$ are the dipole sums taking into account the influence of electromagnetic coupling between the particles in the array on their dipole moments. The expressions for the polarizabilities $\alpha^E$ and $\alpha^M$ obtained from Mie theory, and for $G^0_{xx}$ and $G^0_{yy}$ can be found, for example, in [1] and [13], respectively. From Eq. (1), the intensity transmission coefficient is given by

$$|t|^2 = \left[1 - \frac{k_d}{2S_L}\left[\text{Im}(\alpha^{\text{E}}_{\text{eff}}) + \text{Im}(\alpha^{\text{M}}_{\text{eff}})\right]\right]^2 + \frac{k_d^2}{4S_L^2}\left[\text{Re}(\alpha^{\text{E}}_{\text{eff}}) + \text{Re}(\alpha^{\text{M}}_{\text{eff}})\right]^2. \quad (2)$$

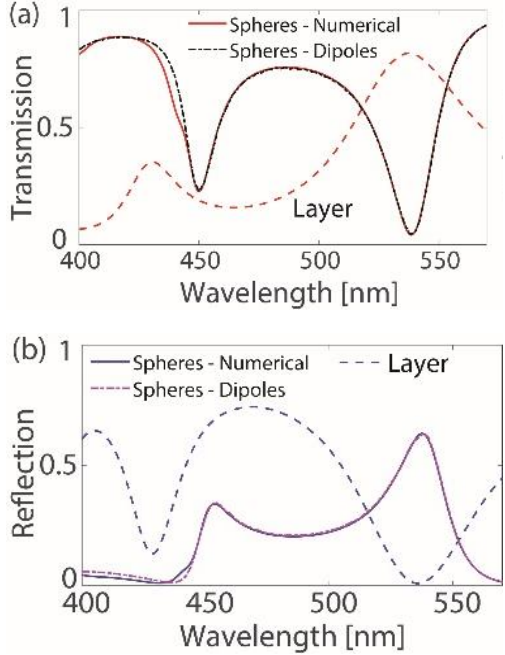

Fig. 1. (a) Transmission and (b) reflection coefficients of continuous Si layer of 130 nm thick and Si nanosphere array with radius $R$ = 65 nm and periodic arrangement in a square array with the period $D$ = 300 nm. The results for the layer are obtained by numerical simulations. The results for spheres obtained by numerical simulations (denoted as "Numerical") and by analytical calculations based on the coupled dipole model and Mie theory (denoted as "Dipoles"). The surrounding medium is air $n = 1$.

Analyzing Eq. (2) we observe that under conditions

(i) $(k_d / 2S_L)\left[\text{Im}(\alpha^{\text{E}}_{\text{eff}}) + \text{Im}(\alpha^{\text{M}}_{\text{eff}})\right] \to 1$ and (3)

(ii) $(k_d / 2S_L)\left[\text{Re}(\alpha^{\text{E}}_{\text{eff}}) + \text{Re}(\alpha^{\text{M}}_{\text{eff}})\right] \to 0$

the transmission is suppressed. One can see that whether both conditions are satisfied depends on the particle dipole polarizabilities and array period. In order to check whether these conditions are satisfied, we consider infinite arrays of spherical Si nanoparticles in the dipole approximation [1]. Figure 2a demonstrates the spectra of the effective ED and MD polarizabilities of Si nanoparticle in arrays with different sizes of the square unit cell and different surrounding conditions. One can see that the dipole-dipole interparticle interaction in the array does not significantly affect the resonant dipole polarizabilities with changes of the array period in the spectral range above diffraction (Fig. 2a). In this case, all Rayleigh anomalies are located on the shorter wavelength side with respect to the resonances of the effective nanoparticle polarizability. The second condition (ii) in Eq. (3) can be satisfied only in the spectral region between EDR and MDR because the sign of the polarizability real part of the both ED and MD are opposite at the spectral range. However, to satisfy the first condition (i) in Eq. (3), the dipole scattering from nanoparticle in the array should be sufficiently strong. From Fig 2a, one can see that the imaginary parts of the ED and MD polarizabilities have maxima at the resonances providing a possibility to satisfy the first condition (i), and quickly drop outside the resonance spectral range. Note that the real parts of the polarizabilities are much smaller than the imaginary parts at the resonance spectral ranges (Fig. 2a). Thus, a significant suppression of light transmission can be expected in spectral regions of the two dipole resonances for arrays with unit cells satisfying the estimation

$$S_L \approx \frac{2\pi \text{Im}(\alpha_R)}{\lambda_R} \quad (4)$$

where $\alpha_R$ is the corresponding effective dipole polarizability at the resonant wavelength $\lambda_R$. Far away from Rayleigh anomalies the dipole-dipole interactions weakly affect the single particle polarizability, and for the estimations, one can use single-particle polarizability. For example, the MDR of individual Si nanoparticle with $R$ = 65 nm is excited at the free-space wavelength $\lambda_R \approx 550$ nm, and the corresponding resonant $\text{Im}(\alpha_R) \approx 0.01 \mu m^3$ [1]. Thus, using Eq. (4) and assumption $S_L = D^2$, one can estimate that the strong suppression of light transmission at the MD resonant wavelength can be obtained in an array composed of such Si nanoparticles with orthogonal periods of $D \lesssim 338$ nm. The transmission suppression in Fig. 1a at the wavelength of $\lambda_R \approx 545$ nm just corresponds to this example.

Further, we perform general numerical simulations of transmission (with CST Microwave Studio), varying the array periodicity in different surrounding conditions. The resulted transmission changes with array period variations are shown in Fig. 2b,c in the case when the Rayleigh anomaly is located on the blue side from the individual EDR and MDR and, as a consequence, does not affect them significantly. From the figures, one can see that there are two narrow bands of the transmission suppression for $D > 200$ nm at $\lambda \approx 450$ nm and $\lambda \approx 545$ nm (Fig. 2b) and $D > 600$ nm at $\lambda \approx 1250$ nm and $\lambda \approx 1600$ nm (Fig. 2c). The multipole analysis of individual nanoparticles [1] proves that these bands correspond to the excitation of EDR and MDR. Note, when the Rayleigh anomaly approaches EDR and MDR, the resonance wavelength experiences small redshifts due to the interparticle coupling. One can see that, in both cases (Fig. 2 bc), increasing the array period, the suppression of light transmission is decreased. For the sparser arrays, the compensation point shifts to ED and MD resonance maxima. For the largest periods, the compensation is not achieved, and transmission is relatively large at that wavelength. Importantly For the small periods, the two suppression bands degenerate in one broadband with the inclusion of the spectral region between the ED and MD resonances, where the real parts of the effective ED and MD polarizabilities have opposite signs. It means that in such closely-packed arrays, the suppression of the transmission is determined by simultaneous ED and MD contributions providing destructive interference between incident and scattered waves.

The situation becomes more complicated in the case when Rayleigh anomaly is realized in spectral ranges of the single particle dipole resonances, and LR can be excited due to corresponding dipole coupling in the arrays. Due to excitation of LR, the effective dipole polarizabilities will have an additional resonance, realized on the red side from the corresponding Rayleigh anomaly, and much stronger than the individual particle resonance [1]. Figure 3a shows the dipole polarizabilities of Si nanoparticles from arrays placed in the surrounding medium with $n = 1.5$. So the effective incident wavelengths in the medium are smaller than in free space, and the Rayleigh anomaly appears at $\lambda_{\text{RA}} = nD$ measured in free space without surrounding medium. From Fig. 3a, one can see that there is the ED-LR for the period $D = 350$ nm. Similar to the individual dipole resonances considered above for the arrays in free space, the excitation of LRs also provides the suppression of the array transmission. If ED- and MD-LRs are excited at different spectral regions, the transmission suppressions are enabled by separate contributions of the resonances (Fig. 3bc). However, with increasing period of the array, the Rayleigh anomaly and the corresponding ED- and MD-LRs are shifted to the red side, but with different spectral rates. This is seen in Fig. 3b,c, where the ED-LR and MD-LR are presented as bright yellow regions below the Rayleigh anomaly, the ED-LR is shifted spectrally faster than MD-LR with the array period increase. Therefore, there is a spectral point where ED-LR and MD-LR overlap: for $D = 380$ nm, the overlap wavelength is $\lambda = 580$ nm (Fig. 3b) and for $D = 1100$ nm, $\lambda = 1780$ nm (Fig. 3c). At the overlap conditions, the lattice Kerker effect (suppression of the total reflection) is realized providing sharp increasing of the transmission for arrays with weak absorption (see Fig. 3b at $\lambda = 600$ nm and $D = 380$ nm and Fig. 3c at $\lambda = 1800$ nm and $D = 1200$ nm). If light absorption plays a noticeable role in array optical properties, the transmission coefficient cannot reach the maximum value equal to unite at the wavelength where lattice Kerker effect is observed (where the reflection is totally suppressed). The resonant lattice Kerker effect has been recently demonstrated and investigated elsewhere [13,14]. At the condition of the resonant Kerker effect, the transmitted light can reach the intensity of the incident wave, however, the phase of the transmitted wave is different [8].

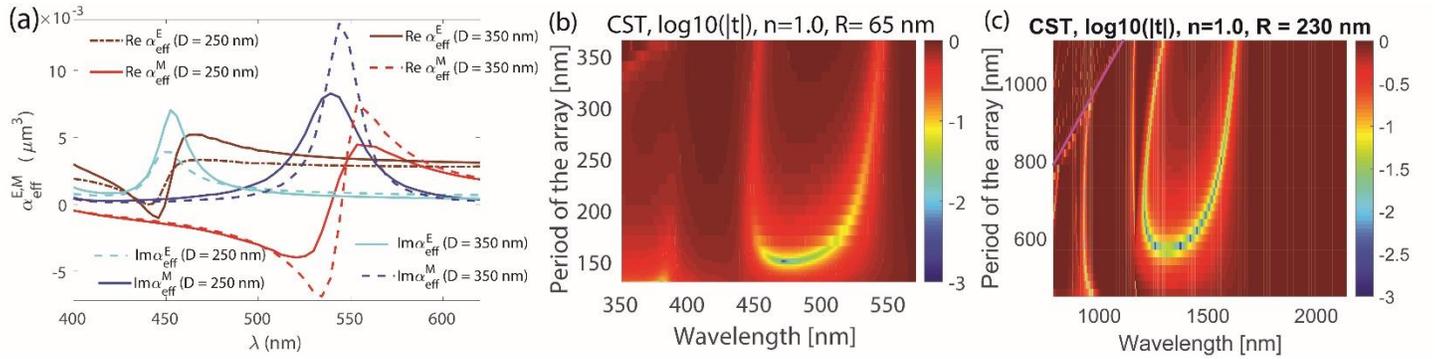

Fig. 2. (a) Effective electric and magnetic polarizabilities of spherical Si nanoparticles with $R = 65$ nm arranged in the infinite arrays with the square elementary cell. The array period $D$ is denoted in the legend. (b),(c) Logarithm presentation of the transmission coefficient of Si nanospheres arrays with square elementary cell calculated for different periods: (b) $R = 65$ nm and (c) $R = 230$ nm. The magenta line marks the Rayleigh anomaly for the first diffractive order. The results in (b) and (c) are obtained by numerical simulations. Colorbar is the same for both panels. The surrounding medium is air $n = 1$ in all three panels.

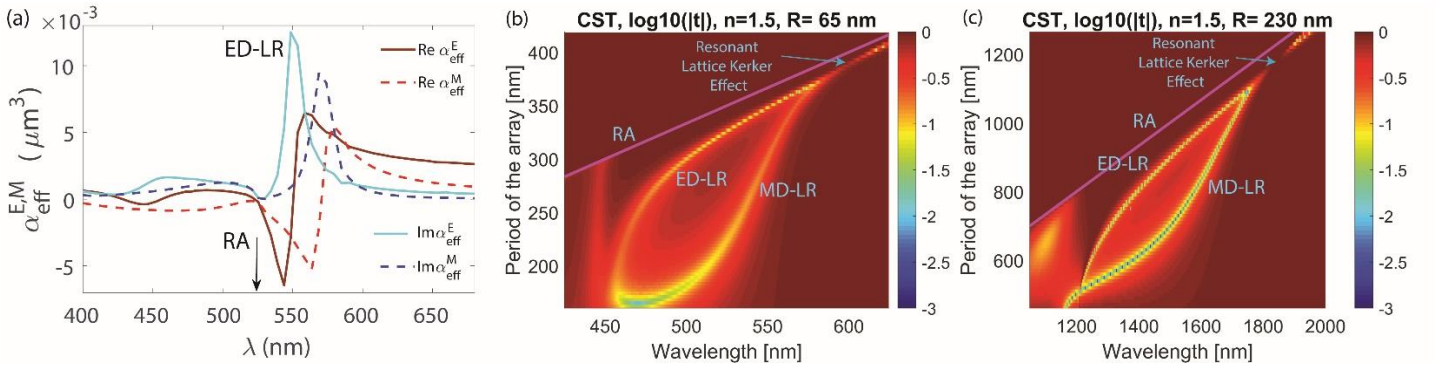

Fig. 3. (a) Effective electric and magnetic polarizabilities of spherical Si nanoparticles with $R = 65$ nm arranged in the infinite arrays with the square elementary cell and period $D = 350$ nm. (b),(c) Logarithm presentation of the transmission coefficient of Si nanospheres arrays with square elementary cell calculated for different periods: (b) $R = 65$ nm and (c) $R = 230$ nm. Magenta lines mark Rayleigh anomaly (RA). The results in (b) and (c) are obtained by numerical simulations. Colorbar is the same for both panels. The data in the diffraction regime (above the Rayleigh anomaly lines in (b) and (c)) is not shown. The surrounding medium has $n = 1.5$ in all three panels, and the wavelengths are presented in free space without the surrounding medium.

General numerical results shown in Fig. 2bc and 3bc are explained in the framework of the dipole model. In order to prove the applicability of the couple dipole approximation, we perform a direct comparison of numerical simulations and analytical model calculations based on Eq. (2). Figures 4 and 5 demonstrate a very good agreement between the analytical calculations and numerical simulations. The agreement is better for larger periods because of the smaller influence of higher multipoles excited inside the particles by incident light. In the near-IR range, silicon losses are negligible, and in this case, the condition of destructive interference is better satisfied, and transmission is almost zero (Fig. 5).

From the calculations in Figs. 4 and 5, one can see that conditions (i) and (ii) of Eq. (3) are satisfied in the vicinity of either EDR or MDR, and either of the resonances can provide a comparable drop in transmission. Thus the analytical model and conditions (i) and (ii) of Eq. (3) allow for an explicit analysis of ED and MD contributions in the transmission suppression effect even for relatively small array periods (Fig. 4a,b where $D = 180$ nm) and can be used for estimations of the spectral points with significant suppression of the transmission.

In conclusion, considering Si nanoparticle structures, we have demonstrated that arrays of high-refractive-index nanoparticles can provide significant suppression of light transmission because of the destructive interference of incident and array-scattered waves. The reported effect is not related to the single-particle or lattice Kerker effects where at least two multipoles are always involved [5,8,13]. In our case, either one or both ED and MD contribute to the significant transmission suppression depending on the array density. It has been explained that the effect is realized due to resonant excitations of ED and MD moments of the nanoparticles including excitation of ED- and MD-LRs. For silicon nanoparticle arrays, the effect is more pronounced in near-IR range than in the visible one because of weak material losses of silicon at higher wavelengths.

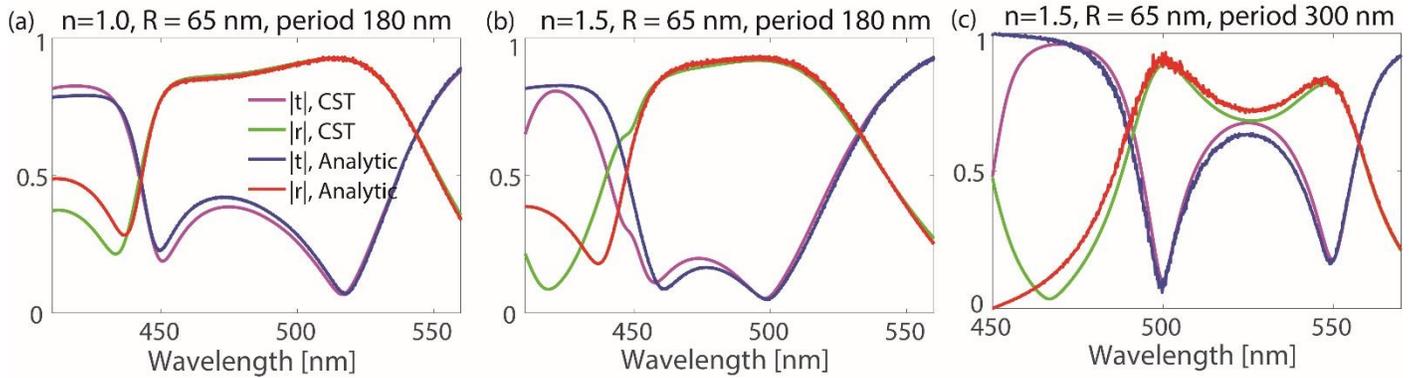

Fig. 4. Transmission and reflection coefficients of Si nanosphere arrays: (a) $D$ = 180 nm, surrounding refractive index $n$ = 1; (b) $D$ = 180 nm, surrounding $n$ = 1.5; and (c) $D$ = 300 nm, surrounding $n$ = 1.5. Sphere radius $R$ = 65 nm and resonances are in the visible spectral range. Legend is the same for all three panels. The plots show a comparison of results obtained by numerical simulations (denoted 'CST') and coupled dipole model (denoted 'Analytic'). In all panels, the wavelengths are presented in free space without the surrounding medium.

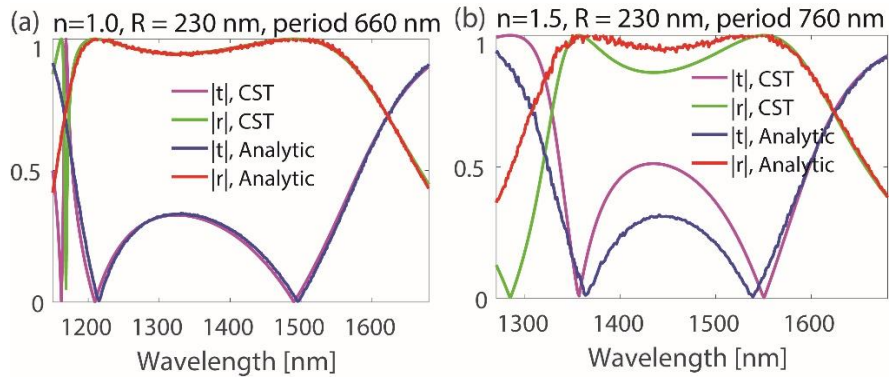

Fig. 5. Transmission and reflection of Si nanosphere arrays: (a) $D$ = 660 nm, surrounding refractive index $n$ = 1, and (b) $D$ = 760 nm, surrounding $n$ = 1.5, with sphere radius $R$ = 230 nm and resonances in the near-IR spectral range (in the proximity to the telecommunication wavelength). The plots show a comparison of results obtained by numerical simulations (denoted 'CST') and coupled dipole model (denoted 'Analytic'). In all panels, the wavelengths are presented in free space without the surrounding medium.

**Acknowledgments.**
The numerical studies have been supported by the Russian Science Foundation (Russian Federation), the project 16-12-10287. This material is based upon work supported by the Air Force Office of Scientific Research under Grant No. FA9550-16-1-0088.